\documentclass[12pt]{elsarticle}
\usepackage{lineno}

\usepackage{natbib}
\usepackage{float}
\usepackage{amssymb}
\usepackage{amsmath}
\usepackage{graphicx}
\usepackage{bm}
\usepackage{epsfig}
\usepackage{multirow}
\usepackage{lineno}
\usepackage{array}
\usepackage{colortbl}
\usepackage{lscape}
\newcommand{\argmin}[1]{\underset{#1}{\operatorname{argmin}}}

\journal{Nuclear Instruments and Methods A}
\begin{document}

\begin{frontmatter}
\title{Parametric unfolding. Method and restrictions. }
\author{Nikolay D.\ Gagunashvili\corref{author}}
\cortext[author] {Corresponding author.\\\textit{E-mail address:}  nikolay@hi.is}
\address{University of Iceland, S\ae mundargata 2, 101 Reykjavik, Iceland}

\begin{abstract}
Parametric unfolding of a true distribution distorted due to finite resolution 
and limited efficiency for the registration of individual events is discussed. 
Details of the computational algorithm of the unfolding procedure are presented. 
\end{abstract}

\end{frontmatter}

\section{Introduction}
The measured distribution $P(x')$ of events with a reconstructed characteristic 
$x'$ obtained from a detector with finite resolution and limited efficiency  
can be represented as
\begin{equation}
   P(x')= \int_{\Omega} p(x)A(x)R(x,x') \,dx,   
\label{p1_main}
\end{equation}
where $p(x)$ is the true density, $A(x)$ the efficiency function describing
the probability of recording an event with a true characteristic $x$, and 
$R(x, x')$ the experimental resolution function, i.e. the probability of 
obtaining $x'$ instead of $x$ after the reconstruction of the event. The 
integration in (\ref{p1_main}) is carried out over the domain $\Omega$ 
of the variable $x$.

If a parametric model of true distribution $p(x; a_{1},a_{2}, \ldots, a_{l})$ 
exists, the model parameters can estimated by fitting
\begin{equation}
 P_s(x') =\int_{\Omega}p(x, a_{1}, a_{2}, \ldots, a_{l}) A(x)R(x,x') \,dx
\end{equation}
to the measured  distribution  $P(x')$, as discussed e.g. in
\cite {zhigunov,zechebook}.

To realize method, the efficiency function $A(x)$ and the resolution function
$R(x,x')$ must be defined. In  many cases, especially in particle physics,  
they are not known analytically and instead are obtained by computer simulation
of the measurement process. A test statistic for comparing the histogram of 
the measured distribution $P(x')$  and the histogram of the measured 
distribution $P_s(x')$ obtained by simulation \cite{gagu_comp} was used 
in \cite{gagu_f}, where the model parameters were estimated by minimization 
of this statistic. The  test  was improved significantly in \cite{chicom}, 
computer code implementing this test was developed in \cite{prev,prev2}.

This paper extends the previous work by presenting a detailed parametric
unfolding algorithm using the above mentioned results \cite{chicom, prev,prev2}. 
A bootstrap algorithm for the calculation of the statistical errors of 
the estimated parameters has been developed. To gauge the quality of the results 
a method of residuals analysis has been developed that complements the $p$-value 
of the chi-square test statistic. Application of the method as well as the 
evaluation are demonstrated on a numerical example.

\section{ Fitting a simulated parametric  model to data}
In experimental particle and nuclear physics analyses the modelling of 
the measurement process usually is the most time-consuming step, requiring 
the simulation of particle transport through a medium and the rather 
complex registration apparatus. The minimization algorithm to estimate
the model parameters then is an iterative procedure that may need many
to calculate the simulated measured histogram many times. A way to 
decrease the CPU-time demand is to perform an initial calculation for some 
distribution $g(x)$, and then to calculate simulated measured histogram  
with an alternative true distribution $p(x; a_{1}, a_{2}, \ldots, a_{l} )$ 
by taking all entries with weights \cite{sobol}
\begin{equation}
  w(x)=p(x; a_{1}, a_{2}, \ldots, a_{l} )/(g(x), \label{weight}
\end{equation}
that exploit the equality 
\begin{equation}
  P_s(x') = \int_{\Omega} p(x, a_{1}, \ldots, a_{l} )A(x)R(x,x') \,dx
          = \int_{\Omega} w(x)g(x)A(x)R(x,x') \,dx.  \nonumber
\end{equation}  

Let us denote the sum  of the entries to the $i$th bin of the measured histogram
by $n_{i}, i=1,...,m$,  and  the sum of the weights of the events in the $i$th bin 
of the simulated measured histogram  by
\begin{equation}
   W_i(a_1,..,a_l)= \sum_{k=1}^{n^w_i}w_{ik}(a_1,..,a_l) ,\quad  
   i=1,..,m,  \label{ffweight}
\end{equation}
where $n^w_i$ is the number of events in bin $i$ and $w_{ik}$ is the weight of 
the $k$th event in the $i$th bin. Events that do not register due to inefficiency 
enter in an overflow bin $m$. The total number of events is denoted by 
$n=\sum_{i=1}^{m} n_i$ and total number of simulated events by $n^w=\sum_{i=1}^{m} n_i^w$. 

Let the values of the parameter $a_1,..,a_l$ be fixed. The hypothesis of homogeneity
that the measured histogram with bin contents $n_i$ and the simulated histogram with
bin contents $W_i$ are drawn from the same parent distribution is probed by 
the test statistic
\begin{equation}
  X^2(\hat p_1,..,\hat p_m )
 \!=\!\left(\frac {1}{n}\sum_i  \frac{ n_i^2}{\hat p_{i}}\right) - n 
 + \left(\frac{1}{n^w}\sum_{i \neq k} \frac{ r_{i}W_{i}^2}{ \hat p_{i}}
 + \frac{(n^w-\!\sum_{i \neq k} r_{i}W_{i})^2}{1-\sum_{i \neq k}r_{i} \hat p_{i}}\right)
 - n^w, \label{stat}
\end{equation}
where the first sum is over all bins $i$, and the second sum omits the least 
sensitive bin $k$ as defined below. The estimates $\hat{p}_1,\ldots,\hat{p}_m$ 
minimize $X^2$, 
\begin{equation}
\hat p_1,..., \hat p_m=\argmin{ p_1,..., p_m} \, X^2(  p_1,...,  p_m),
\end{equation}
subject to the constraints
\begin{equation}
p_i>0\;\forall\,i, \quad  
\sum p_i=1 \quad\mbox{and}\quad  
 1-\sum_{i \neq k} r_{i}  p_i > 0
\quad\mbox{with}\quad
r_i=\frac {W_i} {\sum_{k=1}^{n^w_i} w_{ik}^2}. \label{rat4}
\end{equation}
As shown in \cite{chicom}, the power of the test is optimized by the choice 
\begin{equation}
k= \argmin{i} \frac{\hat{p}_i}{r_i}. 
\end{equation}
Statistic (\ref{stat})  has approximately  a $\chi^2_{m-1}$ distribution 
if the hypothesis of homogeneity is valid \cite{chicom}.

Varying the model paramaters $a_1,a_2,\ldots,a_l$, estimators for best fit 
parameters $\hat a_{1}, \hat a_{2}, \ldots, \hat a_{l}$ are found by 
minimization of the statistic (\ref{stat}),
\begin{equation}
   \hat a_1,...,\hat a_l
 = \argmin{ a_1,..., a_l} \, X^2( \hat p_1,..., \hat p_m,  a_1,...,a_l) \;.
\end{equation}
If the parametric model fits the data, the statistic 
$X^2( \hat p_1,..., \hat p_m, \hat a_1,...,\hat a_l )$ has a 
$\chi^2_{m-1-l}$ distribution, because $l$ parameters are estimated in 
addition to the probabilities and can be used for a goodness-of-fit 
test in the selection of the best from a set of alternative models.

Another approach to the evaluation of the fit quality is the analysis 
of the residuals. The definition of Pearson's residuals for usual 
histograms  is
\begin{equation}
 res_i=\frac{n_i-n\hat{p}_i }{\sqrt{n \hat{p}_i (1-\hat{p}_i)}}.
\end{equation}
which for weighted histograms generalizes to
\begin{equation}
res^w_i=\frac{W_i-n^w \hat{p}_i}{\sqrt{n^w\hat{p}_i(r_i-\hat{p}_i)}}.
\end{equation}
For a homogeneity test two unweighted histograms an adjustment of 
the residual was proposed in \cite{haberman}, which for histograms with weighted entries becomes
\begin{equation}
  Res_i= \frac {res_i}{\sqrt{1-n/(n+q)}}  
\end {equation}
and
\begin{equation}
  Res^w_i= \frac {res^w_i}{\sqrt{1-q/(n+q)}},
\end {equation}
where $q$ is {\it{equivalent number of unweighted events}} for a sample of 
weighted events 
\begin{equation}
q=\frac {(\sum_{i,k} w_{ik})^2}  {\sum_{i,k} w_{ik}^2}.
\end{equation}
If the hypothesis of homogeneity is valid, then the adjusted residuals are 
approximately independent and identically distributed random variables with 
a standard normal PDF $\mathcal{N}(0,1)$.

The statistical errors of parameters can be estimated by the bootstrap 
method \cite{efron}. To realize this method, a set of resampling histograms 
is generated, each according to a multinomial distribution with parameters 
$n, \hat p_1,...,\hat p_m$.  The fit is done for each histogram of the set. 
The resulting set of parameter estimates then permits one to calculate an 
estimate of the covariance matrix of the parameters also.

\section{Numerical example}
Starting from a true PDF to be of the form
\begin{equation}
p(x) =\frac{2}{3\pi}\; \frac{1}{(x-10)^2+1}+\frac{1}{3\pi}\;\frac{1}{(x-14)^2+1}, \label{testform}
\end{equation}
the measured density $P(x')$ was defined according (\ref{p1_main}) with an acceptance 
function
\begin{equation}
 A(x)=1-\frac{(x-10)^2}{36} 
\end{equation}
and a gaussian resolution function 
\begin{equation}
R(x, x')=\frac{1}{\sqrt{2\pi}\sigma}\exp\left(-\frac{(x'-x)^2}{2\sigma^2}\right), \, \sigma=1\;.
\end{equation}
A simulation of 10\,000 events generated according to $P(x')$ was done 
(see algorithm in  \cite {jinst}) and is presented as a histogram with 77 bins 
in Figure 1.

\begin{figure}[t]
\centering
\includegraphics[width=0.8\textwidth]{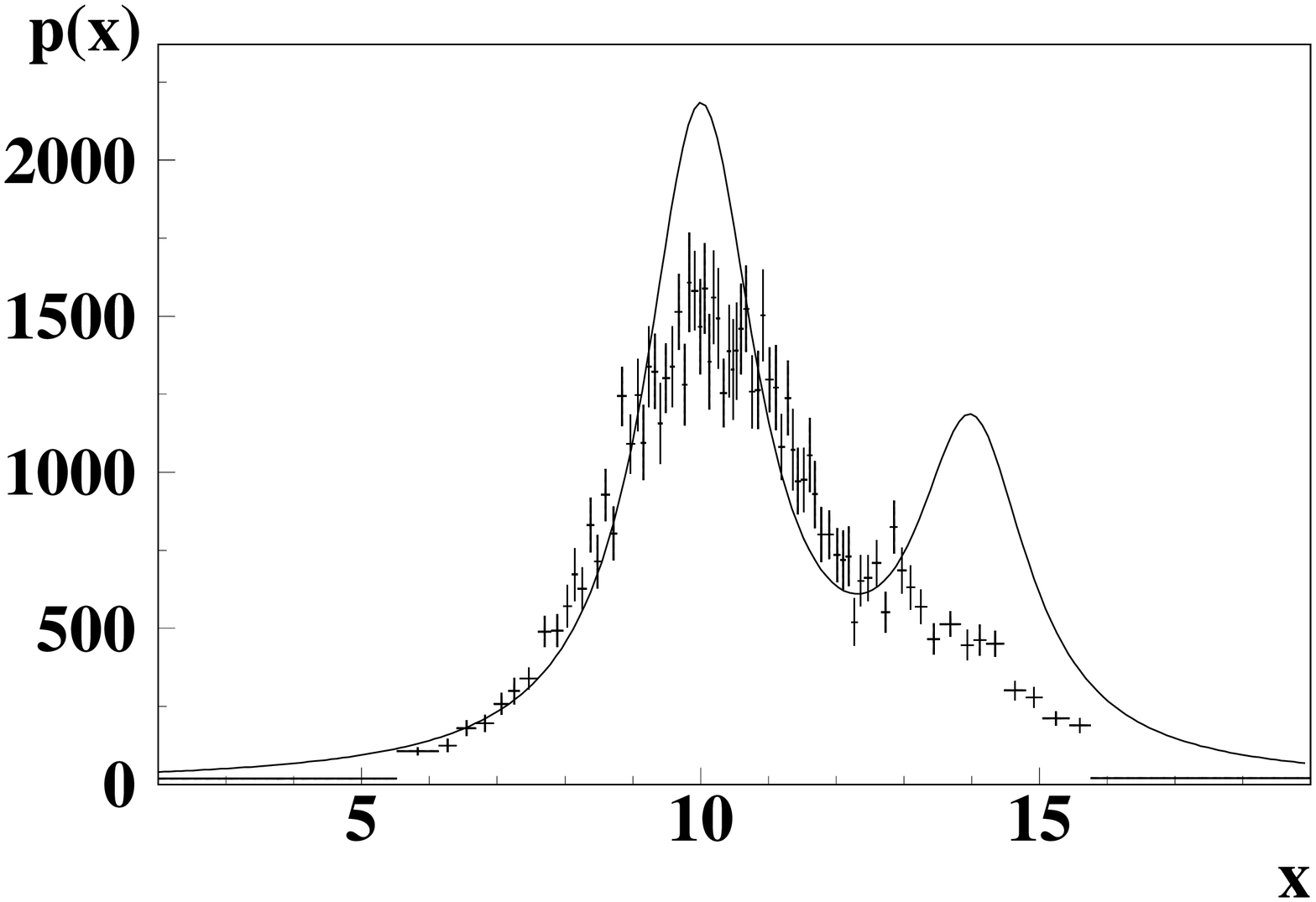}
\caption{Histogram of true distribution $p(x)$ and measured distribution $P(x')$.}
\end{figure}

As input for the fitting procedure the true distribution assumed to be fully simulated 
is taken as
\begin{equation}
g(x) =\frac{2}{7\pi}\; \frac{1}{(x-9)^2+1}+\frac{5}{7\pi} \;\frac{1}{(x-13)^2+1}.  \label{testform}
\end{equation}
The result $P_s(x')$ from simulating 1\,000\,000 events according to the algorithm 
described in \cite {jinst} is shown in Figure 2 together with the initial distribution
$g(x)$.

\begin{figure}[t]
\centering
\includegraphics[width=0.8\textwidth]{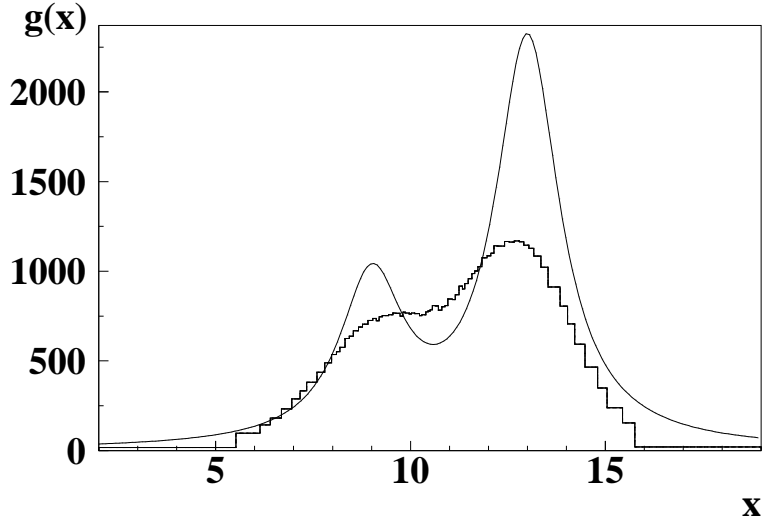}
\caption{Histogram of true distribution $g(x)$  and measured distribution $P_s(x')$.}
\end{figure}

As a fit-model for the true distribution the following function with five free
parameters was chosen,
\begin{equation}
  p(x;a_1,a_2,b_1,b_2,p_0) = \frac{p_0}{b_1\pi}\; 
     \frac{b_1^2}{(x-a_1)^2+b_1^2}+\frac{1-p_0}{b_2\pi}\;\frac{b_2^2}{(x-a_2)^2+b_2^2}.
\end{equation}
In the fit event weights $w(x)=p(x,a_1,a_2,b_1,b_2,p_0)/g(x)$ were used according 
to the formula (\ref{weight}). The test statistic for $X^2(\hat p_1,..,\hat p_{78})$  
for a fixed set of parameters $ a_1,a_2,b_1,b_2,p_0$ was calculated with methods  
and codes published in \cite{chicom, prev, prev2}. The parameter variations were
driven by the SIMPLEX algorithm of the package MINUIT \cite{minuit} in order to 
determine the best fit parameters by minimizing
$X^2( \hat p_1,..., \hat p_{78}, \hat a_1,\hat a_2,\hat b_1,\hat b_2,\hat p_0)$. 
Figure 3 shows unfolded distribution compared to the true PDF. Figure 4 shows 
the weighted histogram with the optimal values of parameters in comparison
with the histogram representing the measured events.

\begin{table}[t]
\begin{center}
 \begin{tabular}{l c c c c c }
$\hat a_1 =10.025\,^{+0.051}_{-0.038}$  &  1 \\
$\hat a_2 =14.007\,^{+0.084}_{-0.086}$  & 0.648 & 1 & \\
$\hat b_1 =~~0.973\,^{+0.061}_{-0.056}$ & 0.648 & 0.459 & 1 & \\
$\hat b_2 =~~0.972\,^{+0.137}_{-0.117}$ &-0.586 &-0.631 &-0.679& 1\\
$\hat p_0 =~~0.667\,^{+0.025}_{-0.021}$ & 0.754 & 0.663 & 0.882&-0.830 & 1\\
&$a_1$&$a_2$&$b_1$&$b_2$& $p_0$
\end{tabular}
\caption{Best fit parameters with uncertainties and correlation matrix.}
\end{center}
\end{table} 
The bootstrap method, with the resample size equal to  1000, was used for estimating  
error intervals and  correlation matrix of the fit parameters. To resample the measured 
histogram, bins contents were generated as multinomial random numbers \cite{cern} with 
parameter $n=10\,000$ and probabilities $\hat p_1,..., \hat p_{78}$.  
Assignment of central value and estimate of the statistical errors of particular parameter, 
for example $a_1$, is done based on the ordered list of bootstrap estimates 
\begin{equation}
\hat a_{1(1)},...,\hat a_{1(1000)},
\end{equation}
where the number in parentheses shows the location when sorting in ascending order.


Starting from the smallast size 68\% confidence interval for $\hat a_1$, with lower and 
upper limit estimated by
\begin{equation}
L= \hat a_{1(\hat i)} \quad\mbox{and}\quad U= \hat a_{1(\hat i+680)},
\end{equation}  
where 
\begin{equation}        
 \hat i=  \argmin{ i} \, [\hat a_{1( i+680)}- \hat a_{1( i)}]    \;
\end{equation}  
the central value is taken to be $ \hat a_{1(\hat i+340)} $ and the uncertainties are 
estimated by the  signed deviations from the central value
\begin{equation}
-err=L-\hat a_{1(\hat i+340)} \quad\mbox{and}\quad +err= U-\hat a_{1(\hat i+340)}.
\end{equation}  

The results are given in Table 1. For the best fit parameter the value of the 
test statistic is 
$X^2( \hat p_1,..., \hat p_{78}, \hat a_1,\hat a_2,\hat b_1,\hat b_2,\hat p_0)=65.310$,
which corresponds to a $p$-value of $p=0.698$. 

A graphical analysis of residuals was done to evaluate the result of then
fitting procedure. Figure 5 shows the distribution of the residuals, 
where a Kolmogorov-Smirnov test of normality gives $p$-value of $p=0.498$. 
Figure 5 shows quantile-quantile plot of residuals with the 95\% confidence 
band \cite{qq}. Figures 3,4,5,6 illustrate the potential of the  
parametric unfolding method, with very satisfactory $p$-values for the 
test statistis considered.
\newpage
\begin{figure}[t]
\vspace *{-2cm}
\begin{center}
\includegraphics[width=0.8\textwidth]{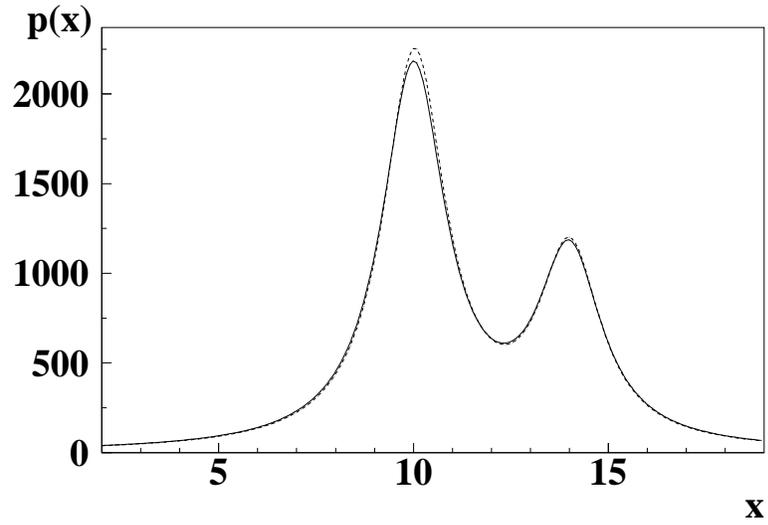}
\caption{ True  PDF $p(x)$  and unfolded PDF $p( \hat a_1,\hat a_2,\hat b_1,\hat b_2,\hat p_0)$ (dashed line)}.
\end{center}
\end{figure}
\begin{figure}[t]
\begin{center}
\includegraphics[width=0.8\textwidth]{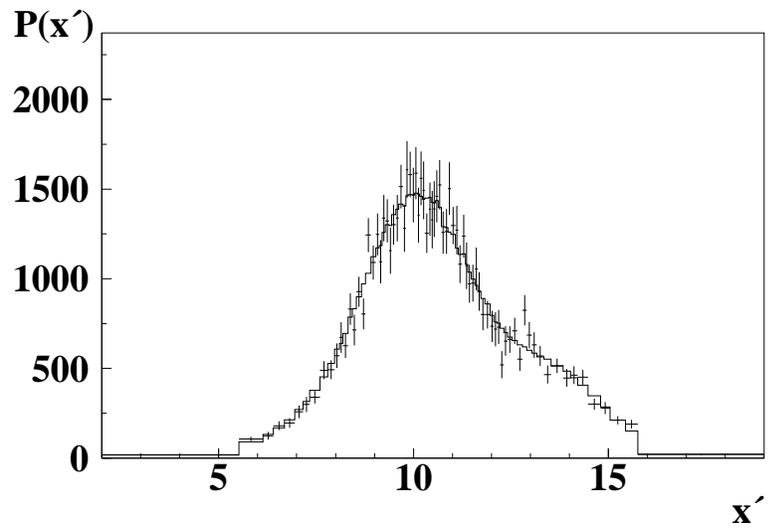}
\caption{Histogram of  measured  PDF $P(x')$  and  histogram  of fitted 
measured PDF  $P_s(x', \hat a_1,\hat a_2,\hat b_1,\hat b_2, \hat p_0)$ 
(solid line)}.
\end{center}
\end{figure}
\begin{figure}[t]
\vspace *{-2cm}
\begin{center}
\includegraphics[width=0.8\textwidth]{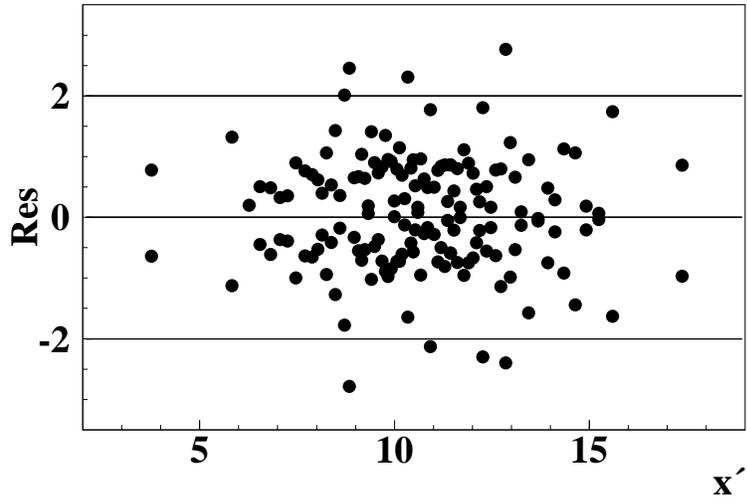}
\caption{Distribution of residuals}
\end{center}
\end{figure}
\begin{figure}[t]
\begin{center}
\includegraphics[width=0.6\textwidth]{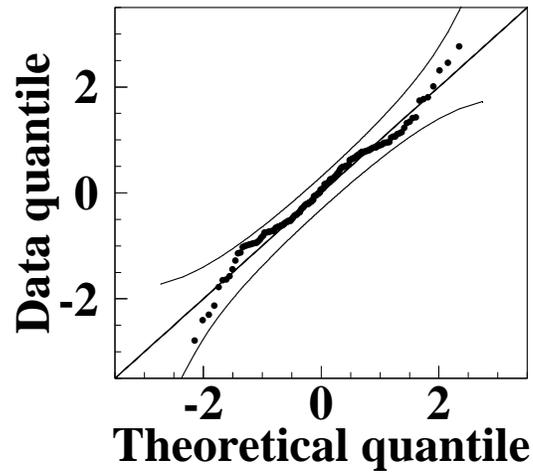}
\caption{Quantile-quantile plot of residuals with 95\% confidence band}.
\end{center}
\end{figure}

\clearpage

\section{Evaluation of the method }
For the evaluation of the method as a whole, the procedure described above
was repeated 1000 times. Sets of 10\,000 events distributed according $p(x)$  
were simulated to create histograms of the measured distribution $P(x')$.
The same set of 1\,000\,000 simulated events distributed according to $g(x)$ 
was used in each run. Figure 7 shows plots for all pairs of the 5 parameter 
and the distribution of the estimators of $\hat a_1,\hat a_2,\hat b_1,\hat b_2,
\hat p_0$. Figure 8 shows the region covered by the 1000 estimates of 
the unfolded distribution together with true distribution $p(x)$. 
Figure 9 presents a histogram of the distribution of the $p$-values and 
confirms that the theoretical distribution $\chi^2_{72} $ can be used for 
a goodness of fit test.   

\begin{figure}[H]
\centering
\includegraphics[width=0.85\textwidth]{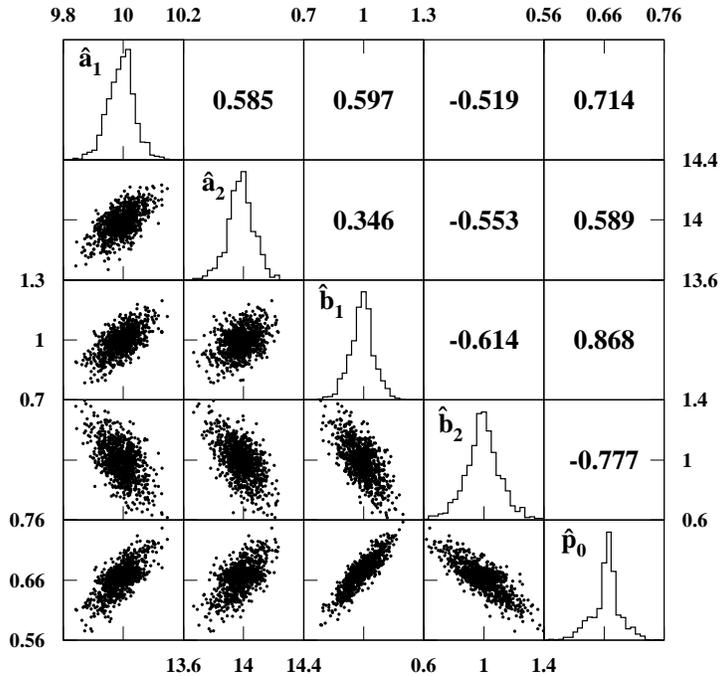}
\caption{Plots for all pairs of estimators of $\hat a_1,\hat a_2,\hat b_1,\hat b_2, 
\hat p_0$ and resulting correlation matrix.}
\end{figure}

\begin{figure}[H]
\vspace *{-2cm}
\begin{center}
\includegraphics[width=0.82\textwidth]{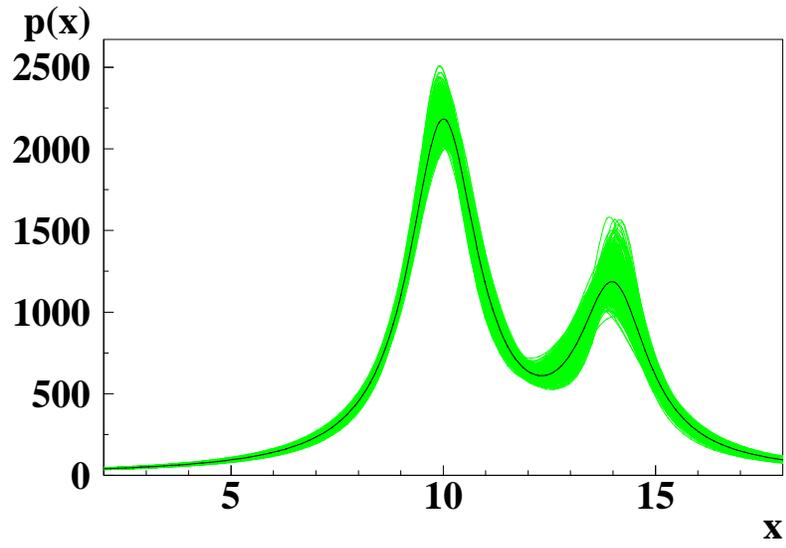}
\caption{Comparison of the regions covered by 1000 estimates of the unfolded 
distribution with true distribution $p(x)$ (solid line) }.
\end{center}
\end{figure}
\begin{figure}[H]
\begin{center}
\includegraphics[width=0.82\textwidth]{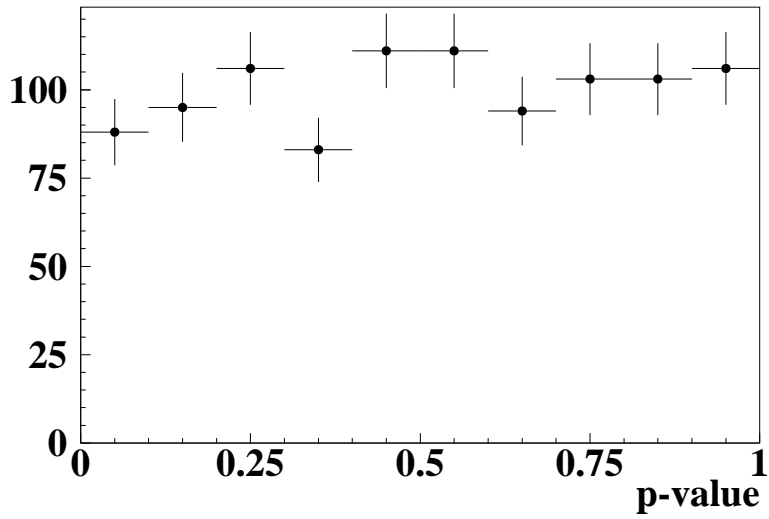}
\caption{Distribution of $p$-values derived from the $X^2$ test statistic.}
\end{center}
\end{figure}




The accuracy of the bootstrap error estimates is checked by doing 1000 toy experiments
and determining the range covered by the smallest  size 68\% quantiles  interval of the parameter values.
The results are given in Table 2, together with the error estimates obtained by
the bootstrap method for the example discussed before. One finds reasonable agreement, 
with some indication that the bootstrap error estimates are slightly conservative. 

\begin{table} [H]
\centering
\begin{tabular}{ l c c c c c }
 \,\,parameter         & \multicolumn{2}{c}{errors }& \multicolumn{2}{c}{errors BS}\\
$a_1=10.000 $          &   +0.036  &   -0.045  &   +0.051  &  -0.038 \\
$a_2=14.000$           &   +0.084 &    -0.076  &  +0.084    & -0.086  \\
$b_1= \,\, \,\,1.000$  &   +0.048 &    -0.062   &    +0.061 &    -0.056  \\
$b_2=  \,\, \,\,1.000$ &   +0.114  &     -0.105&    +0.137 &    -0.117\\
$p_0= \,\, \,\,0.667$  &   +0.017  &   -0.023 &   +0.025   &  -0.021
\end{tabular}
\caption{Error estimates from the smallest size 68\% quantile interval of the parameter 
distributions from 1000 toy experiments compared to the errors obtained 
by the bootstrap method for the example discussed before.}
\end{table} 


\begin{table}[H]
\centering
\begin{tabular}{c c c c c  c}
&$a_1$&$a_2$&$b_1$&$b_2$ &$p_0$\\
$a_1$ & & 0.585 &0.597 &-0.519 & 0.714 \\
$a_2$ &0.648& & 0.346&-0.553&0.589\\
$b_1$&0.648& 0.459& &-0.614&0.868 \\
$b_2$&-0.586&-0.631&-0.679& &-0.777\\
$p_0$&0.754 &0.663 &0.882&-0.830 & 
\end{tabular}
\caption{Correlation matrices obtained by the bootstrap method (lower triangle)  and   
calculated from the distribution of the 1000 simulated toy experiments (upper triangle)}.
\end{table}

Finally, the simulation study was done for different values of the resolution 
parameter $\sigma$. Results are presented in Table 4, showing how the 
accuracy of the parameter estimates diminishes with the worsening of the 
detector resolution. The fits were done without constraints for the values
of the parameters. The study indicates that for larger resolution parameters
eventually constraints will be needed to obtain stable results.

\begin{table}[H]
\centering
\begin{tabular}{ l c c c c  }
\,\,parameter&$ \sigma=0.5 $&$\sigma=0.75$ &$\sigma=1.0$&$\sigma=1.5$\\
$a_1=10.000 $ &  0.049&   0.063 &   0.081 & 0.123 \\
$a_2=14.000$  &  0.096  &  0.122 &   0.160  &0.252  \\
$b_1=  \,\, \,\, 1.000$ &   0.070 &   0.089&    0.110  &0.143  \\
$b_2=  \,\, \,\,1.000$    &   0.147 &   0.179 &    0.219& 0.265 \\
$p_0= \,\, \,\,0.667$   &   0.027 &   0.032 &   0.040 & 0.041
\end{tabular}
\caption{Sizes of 68\% confidence intervals for different values of the 
resolution parameter $\sigma$ in the response function.}
\end{table}

\section*{Conclusions}
Parametric unfolding of data measured by a detector with finite resolution 
and limited efficiency is presented. The method is developed as an application of  
an improved  test for comparing weighted histograms and incorporates new computational  
algorithms and codes. The bootstrap method is employed to estimation the errors of 
the fit parameters. Residual analysis generalized for weighted histograms has been
developed to gauge the quality of the unfolding result. A numerical example is given
to illustrate the method, and an extensive simulation study was done to confirm 
that the proposed method as a whole is valid.

\section *{Acknowledgments}
The author is grateful to Michael Schmeling (Max Plank institute for Nuclear Physics) for critical reading of the manuscript and comments and to Hj\"{o}rleifur Sveinbj\"{o}rnsson and Helmut Neukirchen (University of Iceland) for their help and support of this work. This research was funded by the University of Iceland Research Fund (HI17080029).

\end{document}